\documentstyle[preprint,aps]{revtex}

\begin{document}

\draft

\title{Nonequilibrium Phenomena in Liquid Crystals}

\author{John Bechhoefer}
\address{Dept. of Physics, Simon Fraser University, Burnaby, British
Columbia, V5A 1S6, Canada}
\maketitle
\vspace{1 in}

	The briefest glance through the literature on nonequilibrium
phenomena shows that complex fluids, particularly liquid crystals, are
often favored for experimental investigations.  This might seem
surprising in that complex fluids, as befits their name, are difficult
materials:  experiments require subtle tricks to prepare reproducible
samples; theoretical descriptions lead to notoriously messy equations.
Given the prejudice of physicists towards simple, well-controlled systems,
and given the success that studies of simple fluids have enjoyed, why use
complex fluids to study nonequilibrium phenomena?  In this paper, I shall
offer two answers, one practical and obvious, the other more fundamental and
subtle.

	The obvious, practical reason is that the dynamics of complex fluids
display a variety of interesting ``effects" that have been -- and will
continue to be -- exploited for gain.  Indeed, the use of liquid crystals in
flat-screen displays is perhaps the best-known and most widely exploited of
such special effects.

	The very first observations of liquid crystals
\cite{reinitzer88,lehmann89} noted that although clearly fluid,
they were uniaxially birefringent, a property
that had been associated only with solids.  Even more interestingly, the
optical axis could be aligned along an imposed electric or magnetic field;
soon after, Mauguin \cite{mauguin11} and Grandjean \cite{grandjean16}
discovered
that suitably prepared solid surfaces would also align the optical axes of
nematics.  In the 1930s, Freedericksz and Tsvetkov \cite{freedericksz33} and
Z\"{o}cher \cite{zocher33} put these
two effects together in an experiment illustrated in Fig. \ref{fig:1}.  In
the Freedericksz experiment, the surface and external field tend to align the
molecules in the nematic phase along different directions.  For small
electric field, the orientation imposed by the surfaces wins out; but
above some threshold $E^*$, the molecules align along the imposed
electric field.  The transition is really a
supercritical bifurcation from one uniform state to a second, stationary
state.  This is perhaps the simplest of bifurcations, and one can
immediately expect to see all of the universal behavior associated with such
transitions.  (For example, the maximum deflection of molecules grows as
the square root of the distance above threshold, as suggested in Fig.
\ref{fig:3}a.)

	In 1971, Schadt and Helfrich \cite{schadt71} modified the
Freedericksz experiment slightly by rotating one plate 90${}^{\circ}$ with
respect to the other -- thereby twisting the molecules in the sample -- and
by adding crossed polarizers.  In this configuration, as shown in
Fig.~\ref{fig:2}, the transmitted light-intensity curve
follows that of the molecular
distortion.  The configuration was the basis for the first commercially
successful liquid crystal display and is still extensively used for small
displays where a limited amount of information is to be shown.

	In 1982, it was found that if the twist angle is increased
past 270${}^{\circ}$, the bifurcation becomes subcritical
\cite{waters82,scheffer90}.  (See Fig. \ref{fig:3}.)  Although the resultant
hysteresis causes difficulties for display switching,
the limiting case of a 270${}^{\circ}$ twist angle is useful.
As Fig. \ref{fig:3}c shows, the transmission curves switches more abruptly for
such ``tricritical" bifurcations than for supercritical bifurcations.  (An
elementary analysis shows that the intensity now rises as the distance from
threshold to the 1/4 power \cite{ahlers89}).  This
``supertwist" display is the dominant one used for the large flat-screen
displays found in notebook computers.

	I have outlined the history of liquid-crystal displays in some
detail because -- at least in hindsight -- simple ideas from nonequilibrium
science are relevant.  A good display requires a sharp transition from the
``off" state to the ``on" state.  Thus, it makes sense to use a
supercritical transition, as opposed to a design in which the intensity is
an analytic function of the control parameter.  Changing the
bifurcation from supercritical to sub- or tricritical further speeds the
switching.

	Simple ideas from nonequilibrium science can thus be combined with the
special properties of complex fluids (birefringence, electric-field
alignment of the optical axis) to create useful devices.  The large markets
for such devices -- well over \$3 billion per year for liquid-crystal
displays \cite{nikkei} -- certainly justifies continued research into
understanding and further cataloguing of analogous special effects.
Other special effects I could have cited include drag reduction in
turbulent flows by adding small amounts
of polymer \cite{degennes86}, which has been used to make fire hoses
shoot farther and submarines move faster; the giant swelling transition
in gels \cite{osada93}, which promises robotic ``fingers" that can grasp
delicate parts without damage; and electrorheological
fluids \cite{halsey92}, which are being tested in active automobile
suspensions.  In this conference, K. Amundson \cite{amundson}, R. Larson
\cite{larson}, and H. R. Brand \cite{brand} have discussed other interesting
polymer effects.  I could go on, but I hope the point is clear.

	In addition to the ``bestiary" of special effects, there is a
second, more fundamental reason to study nonequilibrium phenomena in complex
fluids.  Nonequilibrium science can loosely be characterized as the systematic
exploration of systems as some ``stress" is increased.
And, simply put, complex fluids are easier to drive out of equilibrium than
simple ones.

	To understand this remark, consider what I shall call -- with no
disapproval implied~-- the ``conventional" view of the progression of
nonequilibrium phenomena.  This view, largely shaped by work in fluid
dynamics, is sketched in Fig. \ref{fig:4}:  unstressed or lightly stressed
systems are in a simple ``lamellar" state.  As the stress is increased, the
system undergoes a sequence of bifurcations that results in a
time-dependent, chaotic state with limited temporal but full
spatial coherence.  As one further increases the stress, a second series of
transitions -- less well understood -- progressively destroys the spatial
coherence of the system and results in a fully turbulent flow.  Well-studied
examples that illustrate this progression include Rayleigh-B\'{e}nard
convection and Taylor vortex flow \cite{cross}, where ``stress" is
measured by the Rayleigh and the Reynolds numbers, respectively.

	At first glance, the behavior in complex fluids would seem to
parallel that of simple fluids.  For example, when the Freedericksz
experiment is performed on a nematic that tries to align perpendicularly
to the applied field, convective motion is observed.  (See, for example,
W. Zimmerman's contribution to these proceedings \cite{zimmerman}.)  I
want to suggest, though, that there is an important difference between the
behavior of complex and simple fluids when driven out of equilibrium:  In
simple fluids, for reasonable driving stresses, the fluid is always in local
-- but not global -- thermodynamic equilibrium.  For simple fluids, this
observation has a number of consequences.  If, during an experiment on
simple fluids ({\it e. g.}, Rayleigh-B\'{e}nard convection using water),
you were to sample the fluid used, you would find
its material properties to be the same as in equilibrium.  Moreover, at the
end of the day, when you switched off the experiment, the fluid would settle
down to its equilibrium state.  Water that has been churned about at
Reynolds numbers of $10^5$ cannot be distinguished from water that has spent
all day sitting at rest in a glass.  Such observations -- trivial as they
may be~-- stand in contrast to the case of complex fluids where, I shall
argue, modest driving forces can push a system out of equilibrium on length
and time scales comparable to the microscopic scales that characterize the
structure of the fluid.

	Rather than discuss fluid dynamics, I want to focus on a phenomenon
that is equally rich and about which I have personal experience:
solidification.  As is well-known, freezing fronts are often unstable to
shape undulations.  (See Fig. \ref{fig:5}.)  This instability was first
analyzed in detail by Mullins and Sekerka \cite{mullins64} and is relevant
whenever front growth is controlled by diffusive processes (typically, these
are either the diffusion of latent heat or chemical impurities away from the
interface).  If one freezes more rapidly, however, one finds another regime,
the kinetics-limited regime, where front behavior is controlled by local
ordering processes at the interface itself.  As we shall see, the velocity
separating the diffusion- from the kinetics-limited regimes, $v_0$, sets the
scale for nonequilibrium phenomena.  Fronts moving with $v \ll v_0$ are
nearly in equilibrium, while fronts moving with $v \gtrsim v_0$ are strongly
out
of equilibrium.  I shall call the former regime one of slow solidification
and the latter regime one of rapid solidification.

	To understand why $v_0$ sets the scale for nonequilibrium ``stress"
in solidification, we need to recall two facts:  On the one hand, fronts
have a finite thickness $\ell$.  This means that an interface moving at
velocity $v$ will take a time $t_p = \ell /v$ to pass a given observation
point.  On the other hand, a front may be viewed as an ``ordering wave" that
propagates through the fluid.  As the front passes through an observation
point, fluid molecules that were formerly in a disordered state now have to
order.  The ordering takes time -- call it $t_0$.  If the ordering time $t_0
\ll t_p$, then we have slow solidification, since the front has ample time to
order.  If $t_0 \gtrsim t_p$, then the front will have already passed through
the observation point before the ordering is complete, and one may expect
new phenomena to be observed.  Equating the two time scales gives the
velocity $v_0 \sim \ell /t_0$ described above.

	The characteristic solidification speed of a front, $v_0$, is the
ratio of a microscopic length, $\ell$, to a microscopic ordering time,
$t_0$.  For simple fluids, this scale velocity turns out to be roughly the
sound speed, and one can imagine that concocting a controlled experiment
on fronts moving a kilometer a second is not easy!  It turns out, though,
that in a complex fluid, $v_0$ can be dramatically reduced, so that
controlled experiments become feasible.  This is then the second reason that
complex fluids are useful in the exploration of nonequilibrium phenomena.

	To understand where this reduction of $v_0$ comes from, let us first
consider a {\bf simple fluid} -- nice examples include the noble elements,
such as
krypton and xenon --  where the molecules (or atoms) are small and spherical
and where interactions are short-ranged and isotropic.  For such fluids,
the interface width is roughly equal to an atomic diameter, so $\ell \approx
10^{-8}$ cm.  Since all atoms are identical and spherical, the ordering time
is set by the time it takes to remove energy (heat) from the interface.
This is given by the heat diffusion time, so that $t_0 \sim \ell^2/D_h
\approx 10^{-16}$ cm/$10^{-3}$ cm${}^2$/sec $\approx 10^{-13}$ sec.  This
gives $v_0 \sim D_h/\ell \sim 10^{-3}/10^{-8} \sim 10^{+5}$ cm/sec.
($10^3$ m/sec), which is roughly the velocity of sound in a simple fluid.

	Next, consider a {\bf simple alloy}, made of a mixture of two simple
fluids.  The fundamental length scale is still about an angstrom ($\ell
\approx 10^{-8}$ cm), but now the solid phase is formed with an additional
constraint:  not only must energy be removed form the interface, but also the
$A$ and $B$ molecules must be arranged in a precise pattern in the solid
phase.  In addition, the relative concentration of $B$ and $A$ molecules will
differ in the two phases.  Thus, freezing an alloy
requires rearranging atoms, so that the time scale is set
by mass diffusion and not by heat diffusion.  Since the mass diffusivity $D
\approx 10^{-5}$ cm${}^2$/sec is a hundred times smaller than the heat
diffusivity, we expect $t_0 \approx 10^{-11}$ sec. and
$v_0 \approx 10^{-5}/10^{-8} \approx 10^3$
cm/sec (10 m/sec).  Indeed, rapid solidification experiments on metallic
alloys do show interesting phenomena when fronts move faster than about 10
m/sec. \cite{cahn82}

	Notice that the microscopic time scale $t_0$ determining $v_0$ is
set by the slower of the two relaxational processes (heat and mass
diffusion).  This is a general feature of complex fluids: the slowest
relaxational process sets the microscopic ordering time scale.  Notice, too,
that although the length and time scales both increase as we go from a
simple pure fluid to a simple alloy, the ratio $v_0$ decreases.  This, too,
is general.

	Next, we consider {\bf thermotropic liquid crystals}, which are pure
materials made up of rigid, anisotropic molecules.  In most cases, the
molecules are rod-shaped, but disk-shaped molecules also form liquid-crystal
phases \cite{chandrasekhar92}.  The small dimension measures 5 \AA \ across
typically and the large dimension about 30 \AA.  Motions on the scale
of the {\it large} dimension  -- the slowest process -- set the length scale
and the diffusion time
scale.  Although we once again have a pure fluid, the transition from an
isotropic to a nematic state requires orientation alignment, so that one
must consider rotational diffusivities in addition to heat diffusion.  Using
$\ell \approx 10^{-7}$ cm and $D \approx 10^{-7}$ cm${}^2$/sec, we obtain
$t_0 \approx 10^{-7}$ sec and $v_0 \approx 1$ cm/sec.

	{\bf Lyotropic liquid crystal phases} \cite{charvolin89} are formed
when
large amounts of amphiphilic molecules are forced into an aqueous or oily
solvent.  (Amphiphilic molecules have two parts, one hydrophilic, the other
hydrophobic.  Examples include soaps and phospholipids, the constituents of
biomembranes.  See Fig. \ref{fig:5new})
A large variety of phases  -- lamellar, cubic, hexagonal, and others -- can
be observed for different temperatures and amphiphile concentrations.
Here, the repeat distances are larger
($\ell \approx 50 \AA$). Diffusivities vary greatly, ranging from
$10^{-7}$ to $10^{-10}$ cm${}^2$/sec.  The small values occur
because phase transitions can require topology changes in the
amphiphile sheets that are the building blocks of the different
configurations.  Using $D \approx 10^{-8}$ cm${}^2$/sec,
we expect $t_0 \approx 10^{-6}$ sec and $v_0 \approx 1$ mm/sec.

	My final example is the ordering of {\bf diblock copolymers}, which
consist of a chain of $A$ monomers joined covalently to a chain of $B$
monomers \cite{bates90}.  (See Fig. \ref{fig:6}.)  At high temperatures,
the $A$ and $B$
chains are miscible and form a disordered solution.  Below a critical
temperature, the $A$ and $B$ chains phase separate.  In contrast to a
polymer blend, the phase separation must remain local, since the $A$ and $B$
chains remain joined together.  Depending on the relative lengths of $A$ and
$B$ chains, the microscopic ordering will vary.
The phases that are formed have structures similar to those found in
lyotropics.  In
contrast to liquid crystals, polymers -- diblock or normal -- are highly
flexible molecules.  There are new relaxational processes
corresponding to the intricate meshing and disentangling of long polymer
strands.  To estimate this time scale theoretically, we use de Gennes's
``reptation model," in which the polymer molecule is assumed to be confined
to a tube enclosing the molecule \cite{degennes85}.
This gives $t_0 \sim \tau N^3$, where $\tau \approx 10^{-11}$ cm is the
time scale of the monomer (assumed to be a simple
molecule of size $a \sim 10$ \AA).  Alternatively, $t_0$ may be estimated
experimentally from rheological measurements.  (I thank Karl Amundson for
pointing this out.) The appropriate length scale is the
radius of gyration of the molecule, which in a random-walk model is
simply $\ell \sim a N^{1/2}$.  The diffusion constant then is $D \sim
\ell^2/ t_0 \sim (a^2/\tau) N^{-2}$ and the scale velocity for front growth
is $\ell/t_0 \sim (a/\tau) N^{-5/2}$.  Clearly, for large enough $N$, the
velocity scale can be as small as one wishes.  To get reasonable values, one
might want to look at short molecules.  For $N = 150$, we estimate $\ell
\approx 120$ \AA, $t_0 \approx 3 \times 10^{-5}$ sec, $D \approx 4 \times
10^{-8}$ cm${}^2$/sec, and $v \approx 300$ $\mu$/sec.

	The scales for the five examples discussed above are collected in
Table I, where it is immediately clear that increasing the complexity of the
fluid dramatically reduces the velocity scale for rapid solidification.
(For lyotropics, we selected a middle value, $D \approx 10^{-8}$
cm${}^2$/sec, and for the diblocks, we chose $N = 150$.)
Notice that liquid crystals -- both thermotropic and lyotropic -- have
convenient values of $v_0$.  Simple fluids and alloys have $v_0$'s so high
that fronts cannot be followed during an experiment.  Polymers, by
contrast, have scales that are painfully slow, except perhaps for very
short-chained molecules.

	In my own work, I have studied the solidification of thermotropic
liquid crystals with Patrick Oswald, Adam Simon, and Albert
Libchaber \cite{bechhoefer89}.  Our
directional solidification apparatus allowed a maximum speed of about 300
$\mu$m/sec.  This is still somewhat slower than the scale speed of $v_0
\approx 1$ cm/sec, but already interesting phenomena were observed.  In
particular, we observed that in addition to a velocity threshold above which
a flat interface destabilized, there was a second threshold above which the
flat interface reappeared.  In fact, the original study of a flat interface
had predicted that for large freezing velocities and for large thermal
gradients, the front would restabilize.  The front
restabilization velocity is indirectly linked to $v_0$ and occurs at
about 300 $\mu$m/sec for the nematic-isotropic interface of a thermotropic
liquid crystal lightly doped with
ordinary impurities ({\it i.e.,} impurities that are themselves simple
molecules).  A typical stability curve is shown in Fig. \ref{fig:7}.  These
observations were significant in that the restabilization velocity of simple
alloys is on the order of meters/sec.  We were thus able to explore the entire
bifurcation diagram, while previous experiments had probed just a small
piece of it.  We tested the linear stability analysis in the
restabilization regime and also found a number of interesting
secondary instabilities in the interior of the bifurcation diagram (parity
breaking, traveling waves, breathing modes,
etc.). \cite{simon88,flesselles91}

	One answer, then, to the question ``why use liquid crystals and other
complex fluids to study nonequilibrium phenomena" is that they can facilitate
the study of instabilities that were already known in the context of simpler
fluids.  A second answer is that they allow access to the locally
nonequilibrium regime.  What can one expect to see here?  In contrast to the
usual nonlinear regime, much less is known, and I can only suggest what is
to be learned.  If we consider the case of solidification, we see that if
we were to freeze a liquid instantaneously, the disorder of the fluid would
be quenched in and produce a glassy state.  One possibility, then, is that
in the kinetics-limited regime, the ordered state will be progressively
disrupted as the velocity is increased.  The defect density in
the ordered phase would then be a smoothly increasing
function of the freezing velocity \cite{tiller91}.

	Another -- and to my mind, more interesting -- possibility is that
the route from the ordered state of near-equilibrium freezing to the glassy
state of extremely rapid solidification will be marked by a series of
transitions analogous to the phase transitions of equilibrium physics or the
bifurcations of weakly nonlinear dynamics.  With my colleagues Laurette
Tuckerman and Hartmut L\"{o}wen, I have studied a simple theoretical model of
solidification that displays such behavior \cite{bechhoefer91,tuckerman92}.
As illustrated in Fig.
\ref{fig:8}, we have proposed that a rapidly moving front can split into two
separately moving fronts, one dividing the disordered phase (phase 0) from a
new metastable phase (phase 1), the second dividing this metastable phase
from the ordered, thermodynamically stable phase (phase 2).  A necessary
condition for the front to split is that the velocity of the leading edge
$v_{10}$ exceed that of the trailing edge $v_{21}$.  If this condition is
met and if reasonable initial conditions favor splitting, then an
ever-widening region of phase 1 will be created.  Our description of this
transition turns out to be mathematically equivalent to surface melting and
wetting transitions, so that one may view the appearance of phase 1 as being
equivalent to the condensation of a liquid at a solid-vapor interface.
Because the mathematics are the same, one expects to observe a
pretransitional thickening of the 20 interface (logarithmic or power-law
divergence, depending on the nature of the interactions).  In addition, one
can show that the transition can be continuous, hysteretic, or finite
amplitude.  An important difference from, say, surface melting, is that the
transition need not occur in the vicinity of a special point on the
equilibrium phase diagram (for example, the triple point), but can occur
even if phase 1 is metastable at all temperatures.  We require only that its
free energy not greatly exceed that of the stable phase 2 and that
it should somehow ``resemble" the ordered phase.  (For example, one phase
might have an FCC lattice, the other a BCC or simple cubic.)

	Referring to the list of complex fluids in Table I, one might expect
that lyotropic liquid crystals would be good candidates to search for such
behavior.  Not only is the scale velocity $v_0$ modest,
but also lyotropics display a large variety of
phases with weak first-order transitions separating them.  Such experiments
are currently being started in Lyon under Patrick Oswald and at Simon Fraser
University, with Nancy Tamblyn and Anand Yethiraj.  So far, these
transitions have yet to be observed, but the experiments are still
preliminary.

	In the meantime, poor man's versions of the splitting transition
have been observed in thermotropic liquid crystals.  The transition is
not between two thermodynamically distinct phases but between two
configurations of the nematic phase.  In Fig. \ref{fig:9new}, I show a side
view of the meniscus of the nematic-isotropic (NI) interface discussed above.
The glass plates are treated to align surface molecules perpendicular to the
plates (homeotropic orientation).  There is another, globally incompatible
condition at the NI interface itself.  The resulting frustration forces a
singularity in the nematic phase.  (See Fig. \ref{fig:9new}a.)
Topologically, the defect can either be next to the interface or be deep in
the nematic phase.  (See Fig. \ref{fig:9new}b.)  In the latter situation,
the twisted region has a higher elastic energy than the homeotropic region.
The defect line will then move back towards the NI interface at a velocity
$v_{defect}$ set by the nematic's viscosity and elastic constants.  However,
if the isotropic phase is moving faster than the defect line, the defect
cannot catch up and we have the splitting transition described above.  In
this case, the isotropic is phase 0, the homeotropic phase 2, and the new
(planar) orientation of the nematic is the metastable phase 1.  If the
freezing velocity $v$ is low, we expect to see a homeotropic-isotropic
interface (20 interface).  For $v > v_{defect}$, we would expect to see the
defect line peel back, creating a widening region of phase 1.

	In fact, something slightly different happens.  (Fig. \ref{fig:9}.)
The defect line detaches only when $v$ substantially exceeds $v_{defect}$
and then only when the interface passes through a dust particle.  The
interface detaches locally, and a planar region spreads out, creating a
triangular shape that is a record of the space-time history of the new domain.
Note that in Fig. \ref{fig:9} there are simultaneously 20 and 10 interfaces
present.  This means that the splitting transition here is hysteretic.
Finally, while physicists tend to be intrigued by the triangular shape of the
domain, metallurgists are distinctly unimpressed: in the rapid casting of
metal alloys, they see these shapes all the time.

	Summing up, Fig. \ref{fig:10} shows what the complete spectrum of
behavior of a front might be as the driving force is systematically
increased.  In the near-equilibrium regime, the front is unstable to
undulations whose size decreases with velocity.  Above, $v_0$, one can
expect to see front splitting and, eventually, disordering of the
low-temperature phase.  For lack of time, my
discussion of rapid solidification has been incomplete, and I regret not
talking about oscillatory instabilities \cite{karma93} and solute
trapping \cite{aziz82,wheeler}.  Moreover, my focus on solidication was
purely for personal convenience; someone else could have easily rephrased
this talk in terms of the Taylor-Couette experiment, where interesting
features -- including metastable phase formation -- have been
observed for complex fluids undergoing shear.

	I began my discussion by saying that there were two reasons for
using liquid crystals and other complex fluids to study nonequilibrium
phenomena.  The first was that there are a number of special effects that
have great practical application, and I reviewed the history of
liquid-crystal displays by way of illustration.  The second point was the
alteration of microscopic length, time, and velocity scales to values that
are convenient experimentally.  In the end, these two reasons happily do not
separate as neatly as that.  The metastable states that can result from
strongly nonequilibrium processes are themselves new materials,
and they may have useful properties.  Indeed, metallurgists
during the past 30 years have created thousands of new alloys through
rapid solidification, and some of these are widely manufactured.  A very old
example is martensitic steel, which is significantly harder than the
equilibrium austenite steel that is formed at slower cooling rates.  Thus,
although the more fundamentally minded scientist may wish to focus on
strongly nonequilibrium phenomena, the result may
be a better understanding of how to make new materials.

\acknowledgements

	Parts of this work were supported by an AT\&T Bell Laboratories
Graduate fellowship, an NSERC operating grant, and an Alfred P. Sloan
fellowship.  I thank Albert Libchaber, Patrick Oswald, Adam Simon, Laurette
Tuckerman, and Hartmut L\"{o}wen for their contributions to the work on
which this paper was drawn.  I thank Mike Cross for useful comments
about the oral version of this paper.

\begin{figure}
\caption{The Freedericksz transition.  Glass plates are treated to align
molecules in the nematic phase horizontally.  (a)  For small fields, the
molecules lie flat throughout the sample.  (b)  Above a critical field
strength $E^*$, the molecules tilt to align themselves along the electric
field.  The maximum distortion is at the midplane of the sample; the
boundaries still force the molecules to lie flat.}
\label{fig:1}
\end{figure}

\begin{figure}
\caption{The twisted nematic display.  The configuration is similar to that
of the Freedericksz transition, but the bottom plate is rotated 90${}^\circ$,
imposing a twist through the sample.  Crossed polarizers are added to top
and bottom.  (Left.)  With no field applied, the plane of polarization follows
the nematic molecules adiabatically and light is transmitted through the
display.  The changing length of molecules represents rotation out of the
plane of the illustration.  (Right.)  With a field applied, the polarization is
no longer rotated and the analyzer blocks all light transmission.}
\label{fig:2}
\end{figure}

\begin{figure}
\caption{Transmitted light intensity curves for liquid-crystal displays.  (a)
Ordinary twisted nematic cell.  The bifurcation is supercritical; $I(E)$ is
continuous and rises as $(E~-~E^*)^{1/2}$.  (b)  Twist-angle exceeds
270${}^\circ$.  The bifurcation is subcritical, and there is hysteresis in
the switching.  (c) ``Supertwist display," with a twist angle of
270${}^\circ$.  The bifurcation is tricritical, and the intensity increases
above threshold as $(E~-~E^*)^{1/4}$.}
\label{fig:3}
\end{figure}

\begin{figure}
\caption{``Conventional" view of nonlinear phenomena as the driving ``stress"
is increased.}
\label{fig:4}
\end{figure}

\begin{figure}
\caption{A moving nematic-isotropic interface goes unstable as the velocity
is increased.  The interface is on average horizontal, with
the isotropic phase on top and the nematic phase on bottom.  For $v < v^* =$
2.5 $\mu$m/sec, the front is flat.  For $v > v^*$, the front is wavy.  The
bifurcation is supercritical.}
\label{fig:5}
\end{figure}

\begin{figure}
\caption{Sketch of an amphiphilic molecules.  When mixed in high
concentration with water, molecules such as these order in lyotropic liquid
crystalline phases.}
\label{fig:5new}
\end{figure}

\begin{figure}
\caption{Sketch of a diblock copolymer.  These are the polymer equivalent of
amphiphilic molecules and form phases of similar structure to lyotropics.}
\label{fig:6}
\end{figure}

\begin{figure}
\caption{Linear stability of a flat interface under different combinations of
front velocity and the temperature gradient normal to the interface (after
Flesselles {\it et al.})}
\label{fig:7}
\end{figure}

\begin{figure}
\caption{Schematic illustration of a splitting transition.  In (a) and (b),
we plot spatial profiles of a non-conserved order parameter that
distinguishes the two phases.  For example, in a solid-liquid transition,
$q$ could be the amplitude of one of the Fourier amplitudes of a reciprocal
lattice vector.  It is non-zero in the solid but zero in an isotropic
fluid.  For low velocities, the front between phases 2 and 0 propagates
normally.  For high velocities, the 20 front splits into a 21 and 10 fronts.
The 10 front moves faster than the 21 front, leaving a widening region of
the new metastable phase 1.  The dependence of the free energy $f$ on the
order parameter $q$ is sketched at right.}
\label{fig:8}
\end{figure}

\begin{figure}
\caption{Side view of the nematic-isotropic meniscus spanning the gap between
two plates. (a)  Conflicting boundary conditions at the sidewalls and at the
NI interface imply frustration in the nematic, leading
to a defect (disclination line, denoted by the large black dot) in the nematic
phase (here denoted by ``H" for homeotropic orientation).  (b)  The defect
may detach from the interface, creating a region of planar-oriented nematic
(denoted ``P").}
\label{fig:9new}
\end{figure}

\begin{figure}
\caption{Front splitting in a moving nematic-isotropic interface.  The
isotropic phase (phase 0) is on top.  The homeotropically oriented nematic
(phase 2) is the on the bottom.  A region of metastable planar nematic
(phase 1) is present inside the bright triangle.  The simultaneous presence
of 20 and 10 interfaces indicates that the splitting transition here is
hysteretic.}
\label{fig:9}
\end{figure}

\begin{figure}
\caption{A ``different" view of front behavior.}
\label{fig:10}
\end{figure}

\begin{table}
\caption{Microscopic scales of simple and complex fluids}
\begin{tabular}{l c c c c}

 & length scale & diffusion constant & time scale & velocity scale \\
\tableline
 & $\ell$ (cm) & $D$  (cm${}^2$/sec) & $t_0 = \ell^2/D$ (sec) & $v_0 =
\ell/t_0$ (cm/sec) \\
\tableline
simple fluid & $10^{-8}$ & $10^{-3}$ & $10^{-13}$ & $10^5$ \\
binary alloy & $10^{-8}$ & $10^{-5}$ & $10^{-11}$ & $10^3$ \\
thermotropic liq. cryst. & $10^{-7}$ & $10^{-7}$ & $10^{-7}$ & 1 \\
lyotropic liq. cryst. & $10^{-7}$ & $10^{-8}$ & $10^{-6}$ & $10^{-1}$ \\
diblock copolymer & $10^{-6}$ & $10^{-8}$ & $10^{-5}$ & $10^{-2}$ \\

\end{tabular}
\end{table}

\end{document}